\documentclass[review,number,sort&compress]{elsarticle}

\usepackage{subcaption}
\usepackage{graphicx}
\usepackage{amssymb}
\usepackage{color,soul}
\usepackage{wasysym}
\usepackage{url}
\usepackage{ulem}
\usepackage{subcaption}
\usepackage[figuresright]{rotating}

\definecolor{orange}{rgb}{1,0.5,0}
\definecolor{cadmiumgreen}{rgb}{0.0, 0.42, 0.24}

\journal{Applied Radiation and Isotopes}

\begin{document}

\begin{frontmatter}


\title{Tagging fast neutrons from a $^{252}$Cf fission-fragment source}


\author[lund,ess]{J.~Scherzinger}
\author[glasgow]{R.~Al~Jebali}
\author[glasgow]{J.R.M.~Annand}
\author[lund,ess]{K.G.~Fissum\corref{cor1}}
\ead{kevin.fissum@nuclear.lu.se}
\author[ess,midswe]{R.~Hall-Wilton}
\author[lund]{N.~Mauritzson}
\author[lund]{F.~Messi}
\author[lund,ess]{H.~Perrey}
\author[lund]{E.~Rofors}

\address[lund]{Division of Nuclear Physics, Lund University, SE-221 00 Lund, Sweden}
\address[ess]{Detector Group, European Spallation Source ESS AB, SE-221 00 Lund, Sweden}
\address[glasgow]{University of Glasgow, Glasgow G12 8QQ, Scotland, UK}
\address[midswe]{Mid-Sweden University, SE-851 70 Sundsvall, Sweden}

\cortext[cor1]{Corresponding author. Telephone:  +46 46 222 9677; Fax:  +46 46 222 4709}

\begin{abstract}
Coincidence and time-of-flight measurement techniques are employed to tag
fission neutrons emitted from a $^{252}$Cf source sealed on one side with
a very thin layer of Au. The source is positioned within a gaseous $^{4}$He 
scintillator detector. Together with $\alpha$ particles, both light and 
heavy fission fragments pass through the thin layer of Au and are detected. 
The fragments enable the corresponding fission neutrons, which are detected 
in a NE-213 liquid-scintillator detector, to be tagged. The resulting 
continuous polychromatic beam of tagged neutrons has an energy dependence 
that agrees qualitatively with expectations. We anticipate that this 
technique will provide a cost-effective means for the characterization of 
neutron-detector efficiency in the energy range 1 -- 6~MeV.
\end{abstract}


\end{frontmatter}

\section{Introduction}
\label{section:introduction}

We recently reported on our efforts to ``tag" fast neutrons from an
$^{241}$Am/$^{9}$Be source~\cite{scherzinger15} as the first step towards 
the development of a source-based fast-neutron irradiation facility. Here, 
we report on our investigation of a $^{252}$Cf fission-fragment 
fast-neutron tagging technique very similar to that reported on by 
Reiter~et~al.~\cite{reiter06}. In contrast to Reiter~et~al. who employed
a thin layer of plastic scintillator to detect the fragments, we use a 
gaseous $^{4}$He-based scintillator detector. The corresponding fission 
neutrons are detected in a NE-213~\cite{ne213} liquid-scintillator detector. 
This effort represents our first step towards the development of an 
apparatus for the measurement of absolute neutron-detection efficiency at 
our facility.

\section{Apparatus}
\label{section:apparatus}

\subsection{Californium fission-fragment source}
\label{subsection:ff_source}

$^{252}$Cf is an intense source of fast neutrons. With an overall half life 
of 2.645~years and a specific activity of 0.536 mCi/$\mu$g, it decays by 
both $\alpha$-particle emission (96.908\%) and spontaneous fission 
(3.092\%)~\cite{nudat}. The weighted average $\alpha$-particle energy 
is $\sim$6111.69~keV. The prompt-neutron yield is $\sim$3.75 neutrons per 
fission event~\cite{boldeman85,axton85}. The resulting fast-neutron 
energy spectrum follows the Watt distribution~\cite{froner90} and is very 
well known, with a most-probable energy of 0.7~MeV and an average energy of 
2.1~MeV. Our californium source~\cite{cf252} has an active diameter of 5~mm 
and is mounted a capsule that has a thick platinum-clad nickel backside and 
a thin 50~$\mu$g/cm$^2$ sputtered-gold front side which allows both $\alpha$ 
particles and fission fragments to escape.  The (nominal) activity is 
3.7~MBq~\cite{ez_spec}. While trace activity comes from $^{249}$Cf ($<$0.2\%) 
and $^{251}$Cf ($<$0.04\%), the majority comes from $^{250}$Cf ($\sim$7.5\%) 
and $^{252}$Cf ($\sim$92.3\%). We estimate a neutron emission rate of 
$\sim$4~$\times$~10$^{5}$ neutrons per second.

\subsection{Gaseous $^{4}$He fission-fragment detector}
\label{subsection:ffd}

\begin{figure} 
\begin{center}
\resizebox{0.7\textwidth}{!}{\includegraphics{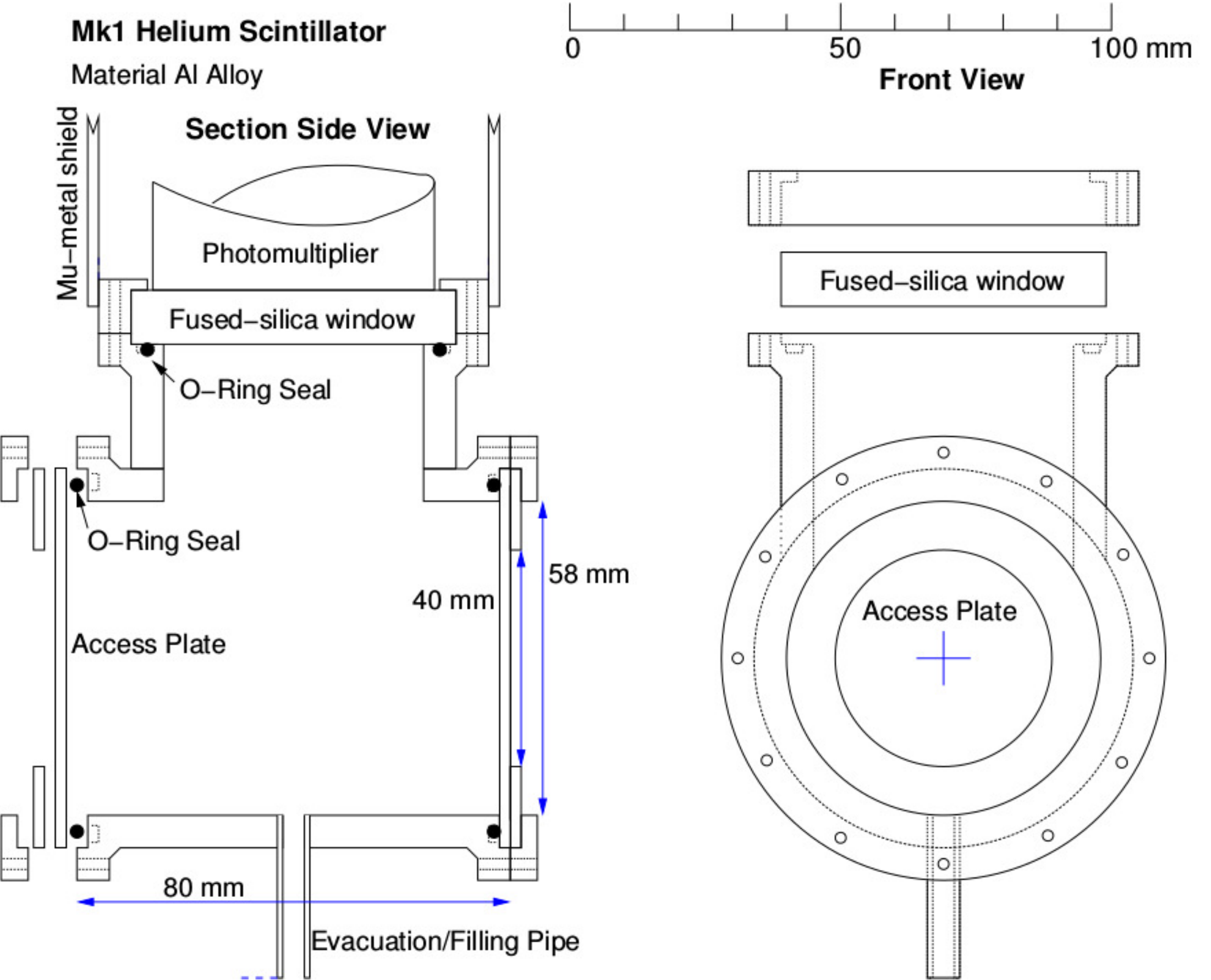}}
\caption{\label{figure:cell_drawing}
A drawing of the gas cell. Left panel: Section Side View. The cell is 
filled with 5 bar of gaseous $^{4}$He scintillator and 500~ppm gaseous 
N$_2$ wavelength shifter. Right panel: Front View. The photomultipier 
tube is attached via a fused-silica window from the top.
}
\end{center}
\end{figure}

The noble gas $^{4}$He is a good scintillator with an ultra-violet light 
yield of about the same magnitude as intrinsic (non Tl-doped) NaI 
crystals~\cite{birks64,dolgosheim69,knoll89,aprile06}. In this measurement, 
we employed a gas cell built originally as a prototype active target for 
recent $^{4}$He photoreaction measurements~\cite{jebali15_2}. The cell was 
machined from a solid aluminum block and has a cylindrical interior volume 
measuring 72~mm long $\times$ 58~mm $\diameter$, for an inner volume of 
$\sim$0.35 liters (see Fig.~\ref{figure:cell_drawing}). 

\begin{figure} 
\begin{center}
    \begin{subfigure}{.30\textwidth}
        \centering
\includegraphics[width=\linewidth]{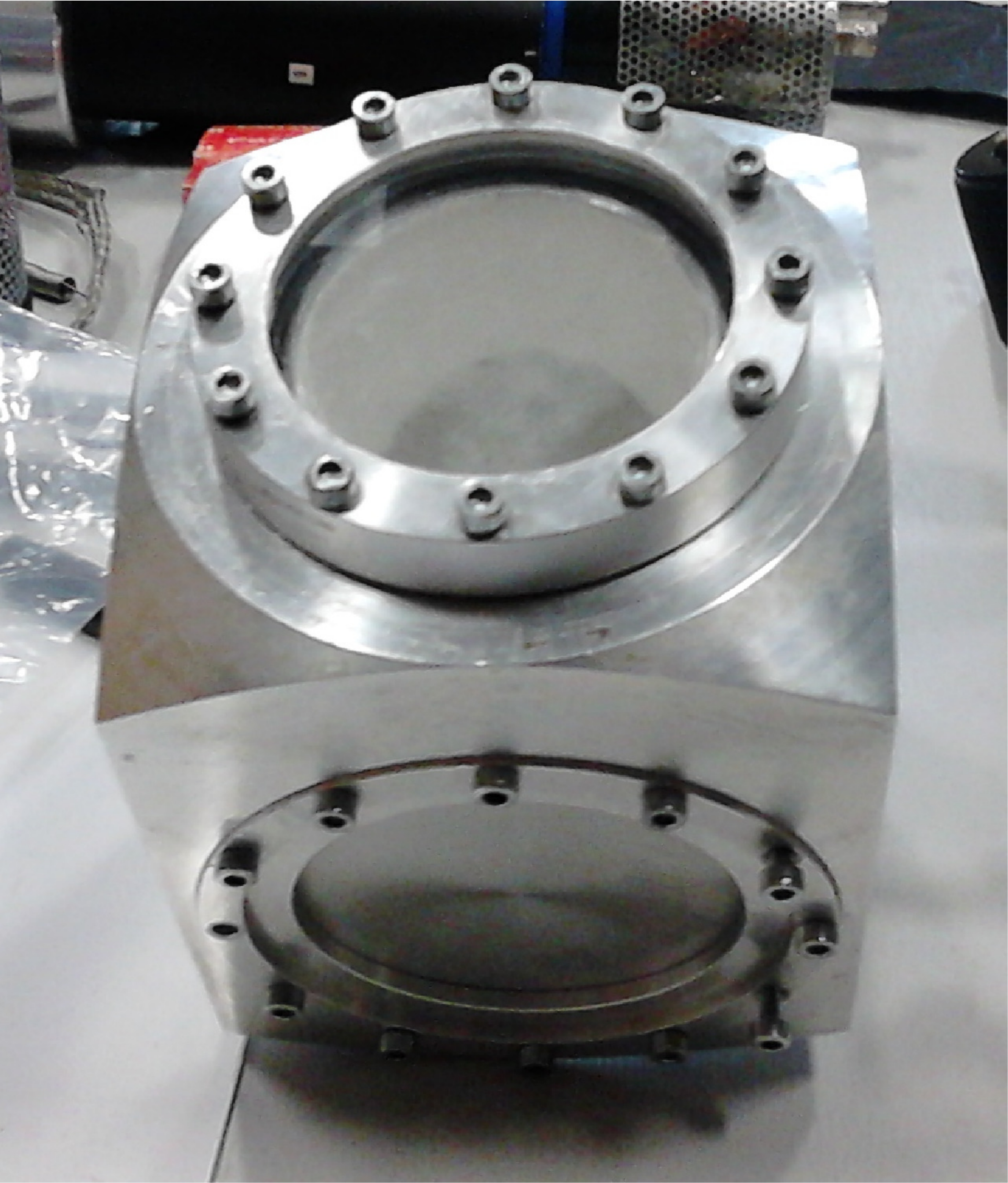}
    \end{subfigure} %
    \begin{subfigure}{.575\textwidth}
        \centering
\includegraphics[width=\linewidth]{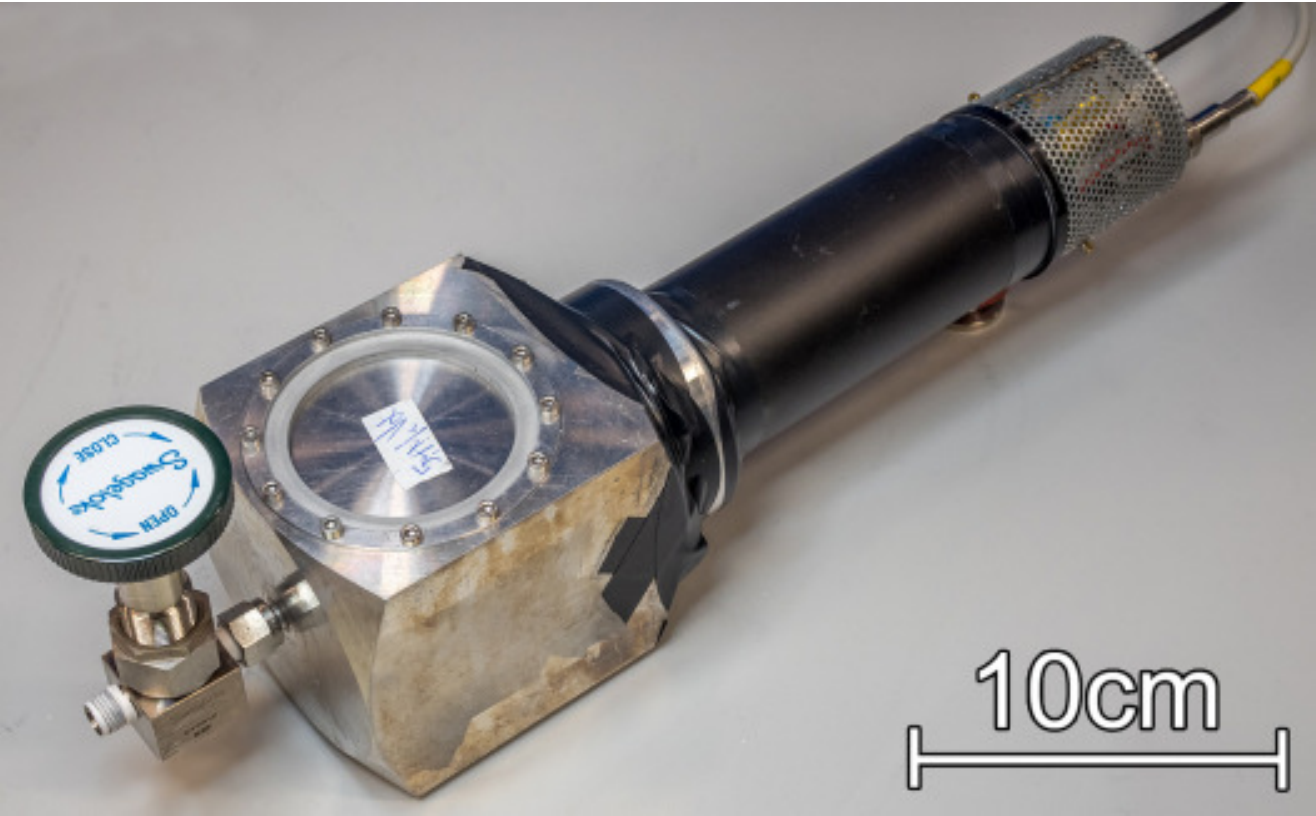}
    \end{subfigure} %
\caption{\label{figure:ff_detector}
Fission-fragment detector. Left panel: photograph of the $^{4}$He gas cell. 
The painted interior of the cell is visible (white). The photomultiplier 
(PMT) tube mounts from the top. Right panel: photograph of the gas cell and 
PMT assembly. The assembly as been rotated and the now-mounted PMT 
corresponds to the black cylinder to the right. 
(For interpretation of the references to color in this figure
caption, the reader is referred to the web version of this
article.)
}
\end{center}
\end{figure}

The interior of the gas cell was sandblasted and then treated with two layers 
of water soluable EJ-510 reflective paint~\cite{ej510}. A fused-silica optical 
window 10~mm thick $\times$ 60~mm $\diameter$ is pressed against the body of 
the cell and allows the scintillation light produced by the $\alpha$ particles
and fission fragments to escape. A rubber O-ring provides the pressure seal. 
The cell was filled 
with 5~bar 99.99999\% pure $^{4}$He (scintillator) gas together with 2.5~mbar 
99.99999\% pure N$_2$ (scintillation-wavelength shifter) gas. A photograph of 
the assembled cell is shown in the left panel of Fig.~\ref{figure:ff_detector}. 
A 5.08~cm XP2262B photomultiplier tube (PMT)~\cite{xp2262} was attached to the 
optical window and EJ-550 optical grease~\cite{ej550} was employed at the 
boundary. A photograph of the assembled detector (gas cell and PMT) is shown 
in the right panel of Fig.~\ref{figure:ff_detector}. 

\begin{figure} 
\begin{center}
\resizebox{0.70\textwidth}{!}{\includegraphics{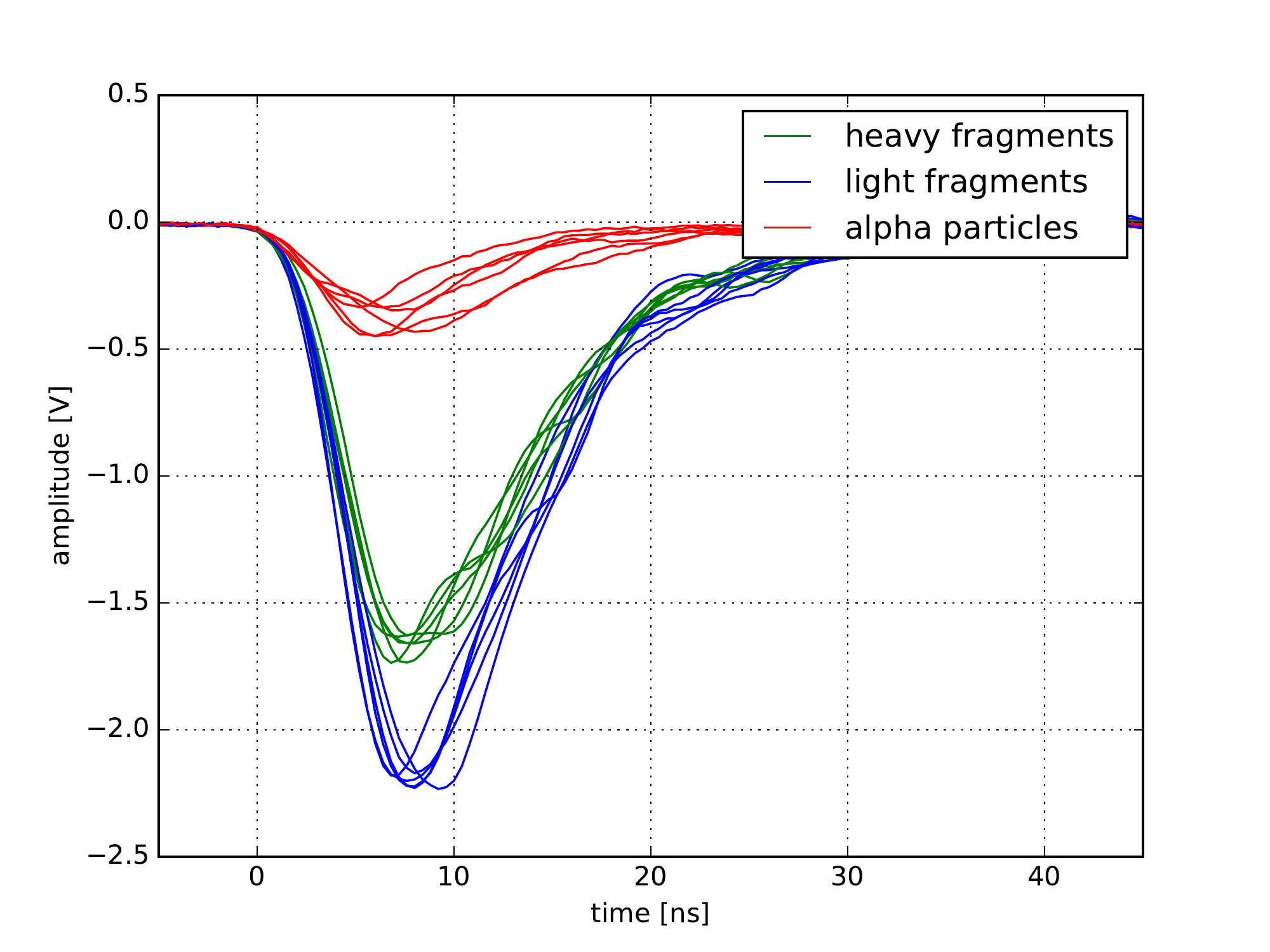}}
\caption{\label{figure:scope_trace}
Typical pulses associated with the decay of $^{252}$Cf obtained with the 
$^{4}$He gas cell. Top (red) trace: $\alpha$ particles. Middle (green) trace: 
heavy fission fragments. Bottom (blue) trace: light fission fragments. (For 
interpretation of the references to color in this figure caption, the reader 
is referred to the web version of this article.)
}
\end{center}
\end{figure}

The californium source described above was positioned at the center of the 
gas cell so that the thin front side through which the $\alpha$ particles 
and fission fragments could escape faced away from the PMT. The distance 
from the californium source to the center of the fused-silica optical window 
was $\sim$65~mm. Operating voltage for the PMT was $-$1750~V and the 
discriminator threshold was set at $-$60~mV. Typical signal risetime was 
5~ns, while the falltime to $<$10\% of the original amplitude was $\sim$10~ns. 
Figure~\ref{figure:scope_trace} shows some typical detector pulses obtained 
with the $^{4}$He gas cell. The top traces with amplitudes of about $-$350~mV 
correspond to $\alpha$ particles. The middle traces with amplitudes of about 
$-$1600~mV correspond to heavy fission fragments. The bottom traces with 
amplitudes of about $-$2200~mV correspond to light fission fragments. We note 
that the average $\alpha$-particle energy is $\sim$6.1~MeV, while the average 
heavy fission-fragment energy is 80~MeV and the average light fission-fragment 
energy is 104~MeV~\cite{vanaarle94}. See also the histogram presented in 
Fig.~\ref{figure:cell_histogram}.

\subsection{NE-213 fast-neutron and gamma-ray liquid-scintillator detector}
\label{subsection:ne213_detector}

NE-213 is an organic liquid scintillator that has been employed for decades
as a fast-neutron detector. The NE-213 liquid-scintillator detector used here 
has been reported upon earlier~\cite{scherzinger15,jebali15,scherzinger16}. 
It consisted of a 62~mm long $\times$ 94~mm $\diameter$ cylindrical aluminum 
``cup" fitted with a borosilicate glass optical window~\cite{borosilicate}. The 
filled cell was dry-fitted against a cylindrical PMMA UVT lightguide~\cite{pmma}
and coupled to a $\mu$-metal shielded 7.62~cm ET Enterprises 9821KB PMT and 
base~\cite{et_9821kb}. Operating voltage was set at about $-$1900 V, and the 
energy calibration was determined using standard gamma-ray sources together 
with a slightly modified version of the method of Knox and 
Miller~\cite{knox72} as described in Ref.~\cite{scherzinger16}. The detector 
threshold was set at 150~keV electron equivalent (keV$_{ee}$), corresponding 
to a neutron depositing an energy of about 1~MeV.

\begin{figure} 
\begin{center}
\resizebox{0.7\textwidth}{!}{\includegraphics{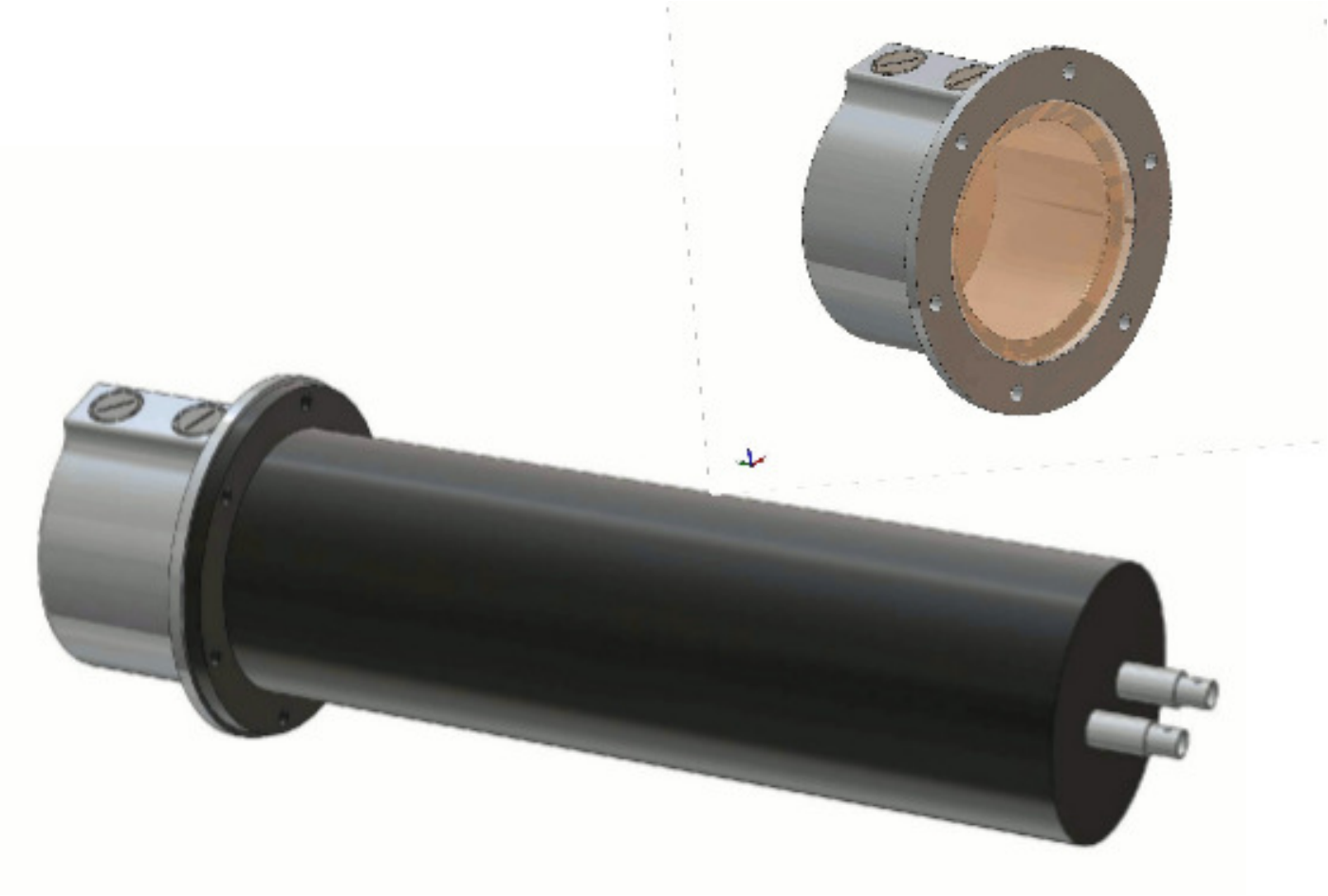}}
\caption{\label{figure:ne213_detector_cad}
The NE-213 detector (CAD representation). Top: the aluminum scintillator 
``cup" which holds the NE-213. A (light brown) window separates the scintillator 
from the lightguide. Bottom: The ``cup" is to the left. The $\mu$-metal 
shielded PMT and base assembly is the black cylinder to the right. Figure 
from Ref.~\cite{jebali15}.
(For interpretation of the references to color in this figure
caption, the reader is referred to the web version of this
article.)
}
\end{center}
\end{figure}

\subsection{Configuration}
\label{subsection:configuration}

\begin{figure} 
\begin{center}
\resizebox{0.75\textwidth}{!}{\includegraphics{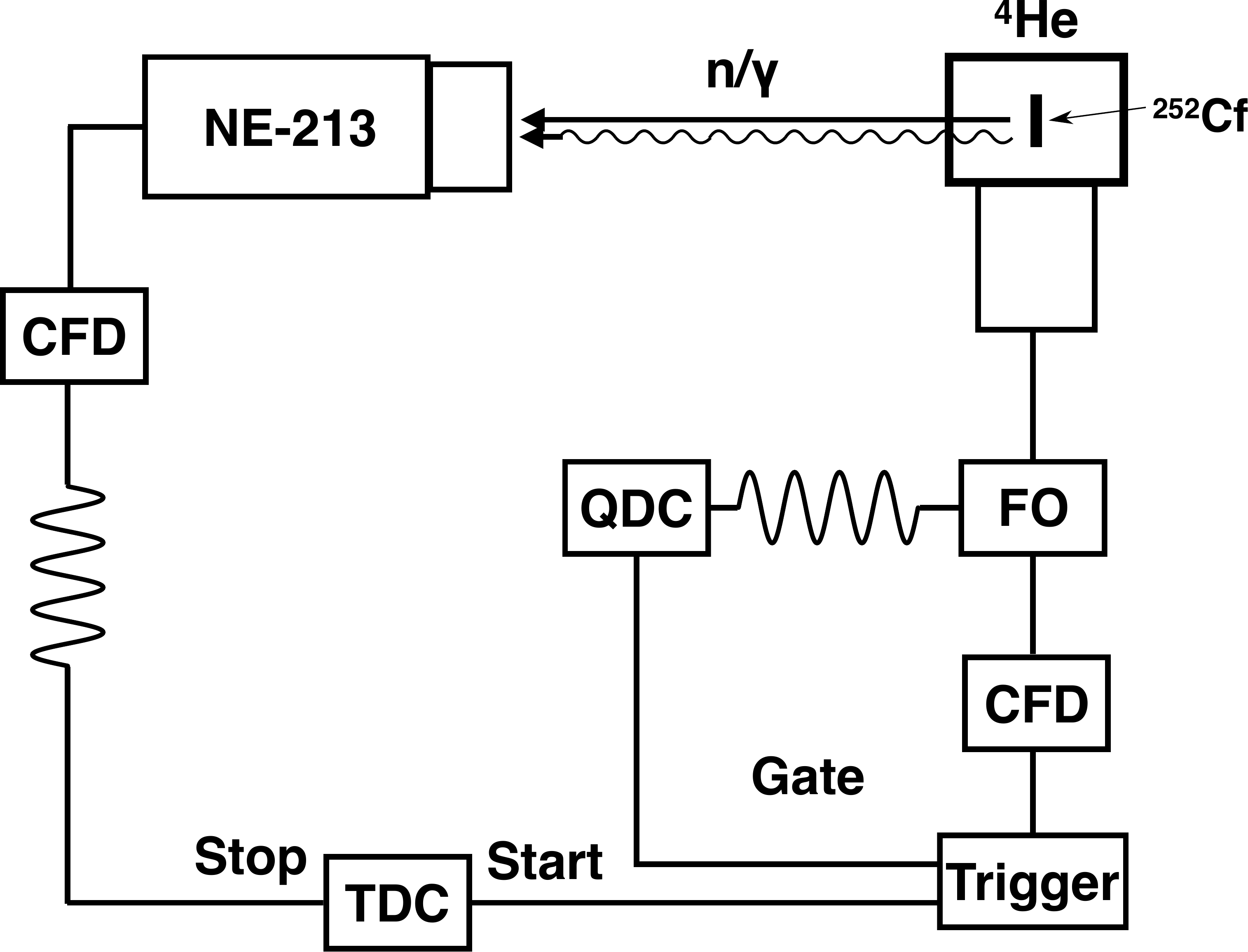}}
\caption{\label{figure:block_diagram}
A simplified overview of the experimental setup (not to scale). The $^4$He 
scintillator detector (which contains the $^{252}$Cf source) and the NE-213 
detector are shown together with a block electronics diagram illustrating the
time-of-flight concept employed.
}
\end{center}
\end{figure}

A block diagram of the electronics is shown in Fig.~\ref{figure:block_diagram}.
$\alpha$ particles and fission fragments were detected in the $^4$He scintillator
detector and corresponding neutrons (and gamma-rays) were detected in the NE-213 
detector. The analog signals from the NE-213 detector were passed to a Phillips 
Scientific (PS) 715 NIM constant-fraction timing discriminator (CFD). The analog 
signals from the $^4$He scintillator detector were fanned out (FO) and passed to 
a PS 715 NIM CFD as well as a CAEN V792 12-bit (DC-coupled 60~ns gate) VME QDC. 
The CFD signals from the $^4$He scintillator detector were used to trigger the 
data-acquisition (DAQ) and thus provided start signals for a CAEN 1190B VME 
multihit time-to-digital converter (TDC) used for the neutron time-of-flight 
(TOF) determination. The NE-213 detector provided the corresponding stop signal.
A SIS 1100/3100 PCI-VME bus adapter was used to connect the VMEbus to a LINUX 
PC-based DAQ system. The signals were recorded and processed using ROOT-based 
software~\cite{root}.

\section{Results}
\label{section:results}

\begin{figure} 
\begin{center}
\hspace*{-1cm}\resizebox{1.20\textwidth}{!}{\includegraphics{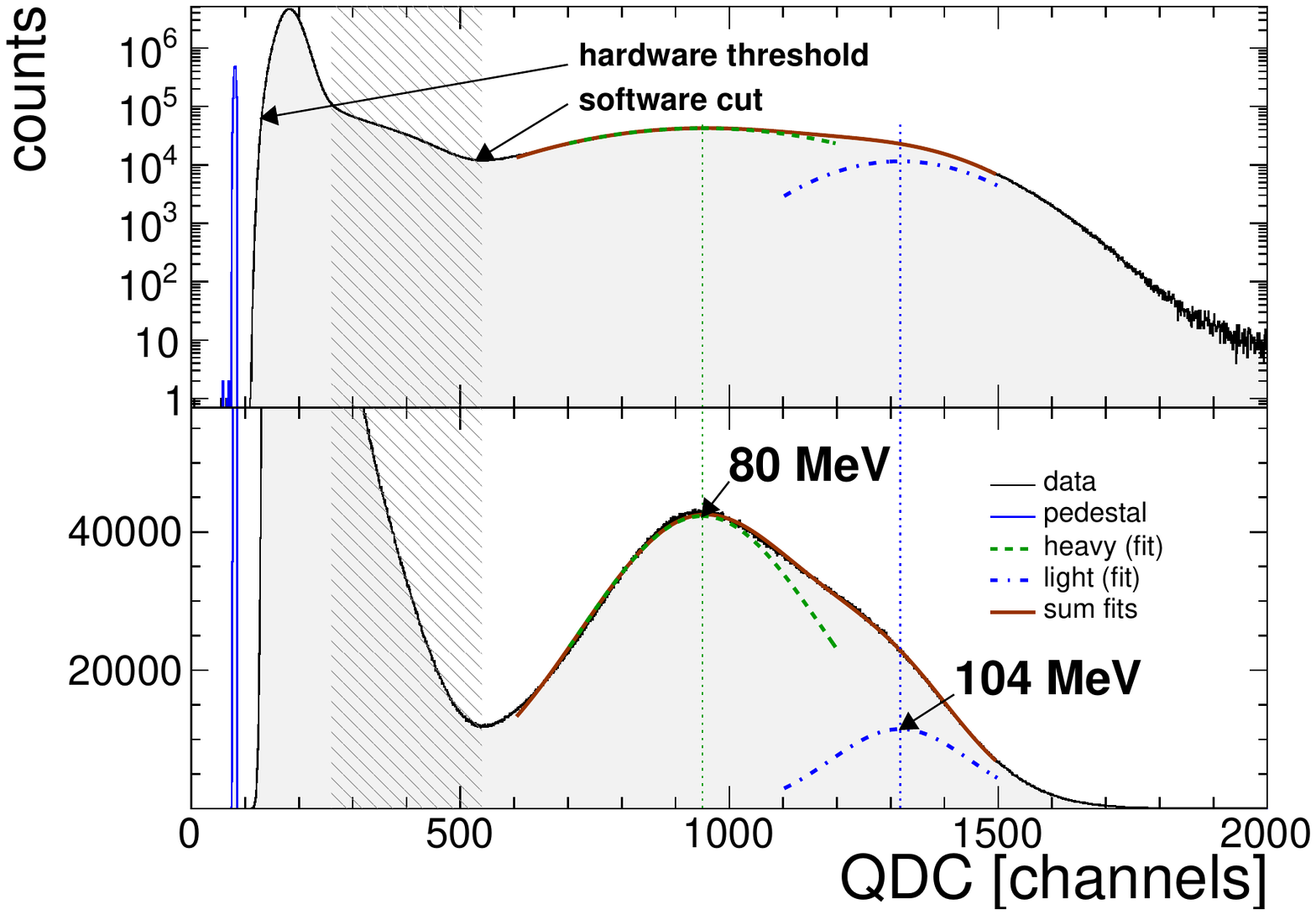}}
\caption{\label{figure:cell_histogram}
Deposited-energy histogram measured using the $^4$He fission-fragment 
detector. Both panels present the same data set. The pedestal (leftmost blue 
peak), hardware discriminator threshold, software fission-fragment cut, and 
distributions corresponding to $\alpha$-particles (dominant peak), heavy 
fission fragments (green dashed line), and light fission fragments (blue 
dot-dashed line) are all shown. 
(For interpretation of the references to color in this figure caption, the 
reader is referred to the web version of this article.)
}
\end{center}
\end{figure}

Figure~\ref{figure:cell_histogram} shows a deposited-energy spectrum measured
using the $^4$He scintillator detector. The top panel is plotted on a 
logarithmic scale to better illustrate the overall features of the spectrum, 
while the bottom panel is plotted on a linear scale to emphasize certain of 
these features. The very sharp leftmost peak in the figure located at about 
channel~80 is the pedestal or zero-energy bin in the QDC. Just to the right 
of the pedestal is the edge corresponding to our hardware threshold located 
at about channel~140. Recall that this discriminator threshold was $-$60~mV.  
The $\alpha$ particles which dominate the spectrum and correspond to the red 
trace in Fig.~\ref{figure:scope_trace} correspond to the peak centered at 
about channel~190. Note that the entire $\alpha$-particle distribution is 
not shown as the hardware threshold cuts into it. A distribution 
corresponding to heavy fission fragments (green trace in 
Fig.~\ref{figure:scope_trace}) is shown here centered at channel~950, while
that corresponding to light fission fragments (blue trace in 
Fig.~\ref{figure:scope_trace}) is centered at channel~1310). Separation of 
fission fragments and $\alpha$ particles is not completely clean, as seen 
in the grey shaded area of Fig.~\ref{figure:cell_histogram} between 
channels~230 and 540. This could result from non-uniform scintillation-light
collection, different energy losses of different particle types in the source
as well as the thin Au source window, non-linearity of the scintillation, or
even fission fragments striking the source holder. This will be examined in 
more detail in a future publication. Recall that the average $\alpha$-particle 
energy is $\sim$6.1~MeV, while the average heavy fission-fragment energy is 
80~MeV and the average light fission-fragment energy is 104~MeV. If we 
calibrate our QDC based upon the average energy deposition of the two types 
of fission fragments and then apply this calibration to the $\alpha$-particle 
distribution, we reconstruct the $\alpha$ peak at $\sim$12~MeV. $^{4}$He is 
often assumed to be a linear scintillator, while this preliminary analysis 
suggests an apparent non-linearity. However, as outlined above, there are 
several factors which will affect the apparent scintillation-pulse height. 
The degree of non-linearity in the scintillation (if any) requires an 
in-depth study. Note that for the data presented subsquently in this paper, 
a software fission-fragment cut located at channel~520 was employed. 

\begin{figure} 
\begin{center}
\resizebox{0.92\textwidth}{!}{\includegraphics{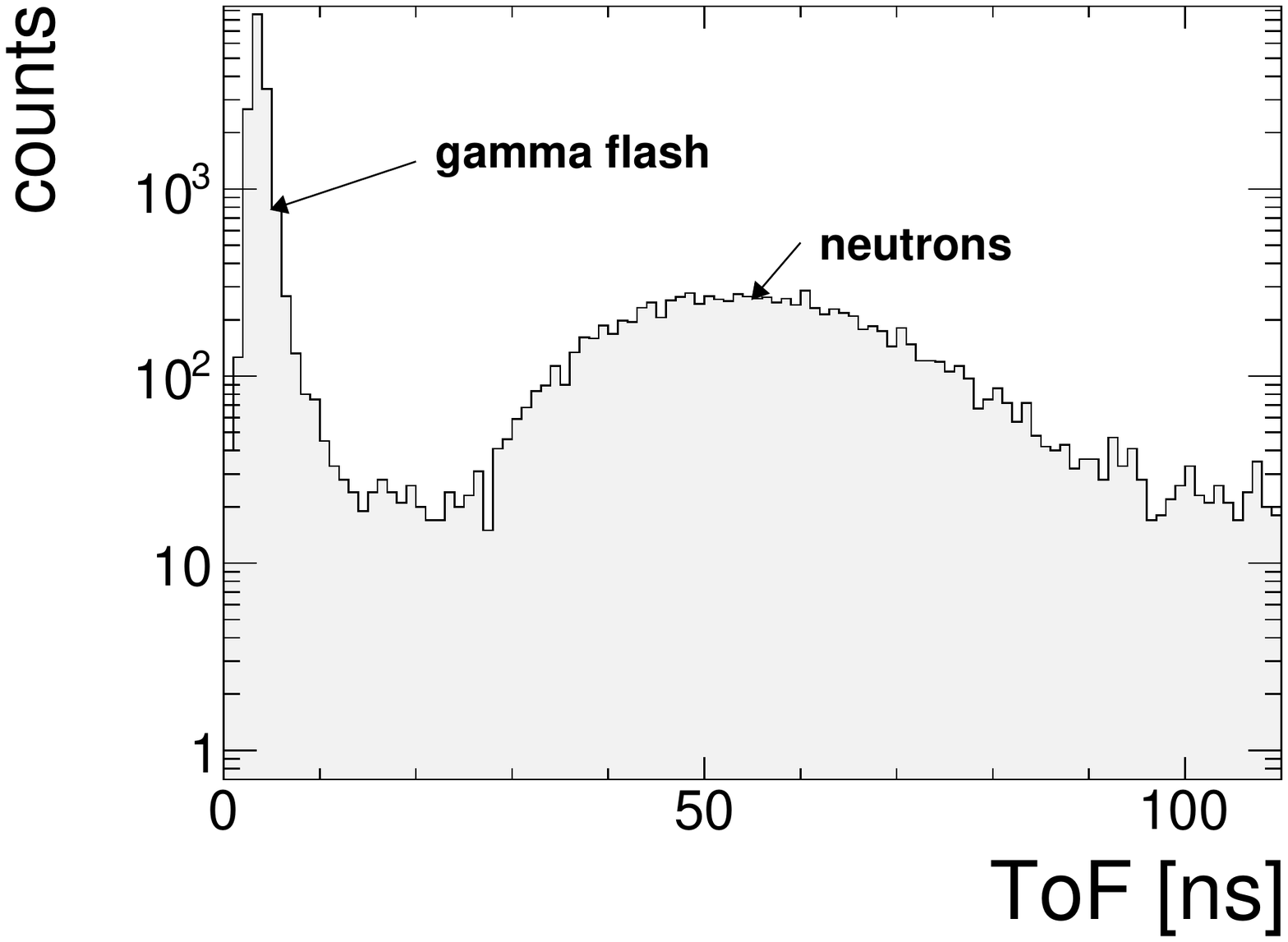}}
\caption{\label{figure:TOF}
Fission-neutron TOF spectrum. The gamma-flash and fission-neutron distribution
are shown. The flat background is due to random events.
}
\end{center}
\end{figure}

Figure~\ref{figure:TOF} shows a fission-neutron TOF spectrum obtained using 
the signal in the $^4$He scintillator detector to start a TDC and a signal 
from the NE-213 detector to stop it. Note that the spectrum shown corresponds 
to events lying above the software fission-fragment cut at channel~520 shown 
in Fig.~\ref{figure:cell_histogram}. After this cut, interpretation of the 
resulting TOF spectrum is straightforward. The sharp peak to the left of the 
spectrum centered at about 5~ns and labeled ``gamma-flash" corresponds to the 
detection of a fission fragment in the $^4$He scintillator detector and a 
correlated fission-event gamma-ray in the NE-213 detector. The $\sim$1.8~ns 
FWHM of the gamma-flash distribution is consistent with the observed timing 
jitter on our PMT signals and the slight tail in the distribution is possibly 
due to time walk in the electronics. Note that $^4$He scintillator is highly 
insensitive to gamma-rays~\cite{jebali15,jebali15_2} and any electrons 
produced via Compton scattering or pair production will result in only a 
very small scintillation signal. These events will be entirely suppressed by 
the relatively high software cut we have applied on the signals from the 
$^4$He scintillator detector. Thus, the present apparatus is almost completely 
insensitive to fission-associated multiple gamma-ray events. The broad bump 
centered at about 55~ns corresponds to the fission-neutron distribution. The 
underlying background distribution corresponds to random coincidences. It was 
measured to be flat as expected by breaking the line-of-sight between the 
$^4$He scintillator detector and the NE-213 detector using a stack of lead 
($\sim$15~cm) and polyethylene ($\sim$10~cm).

\begin{figure} 
\begin{center}
\hspace*{-1.5cm}\resizebox{1.20\textwidth}{!}{\includegraphics{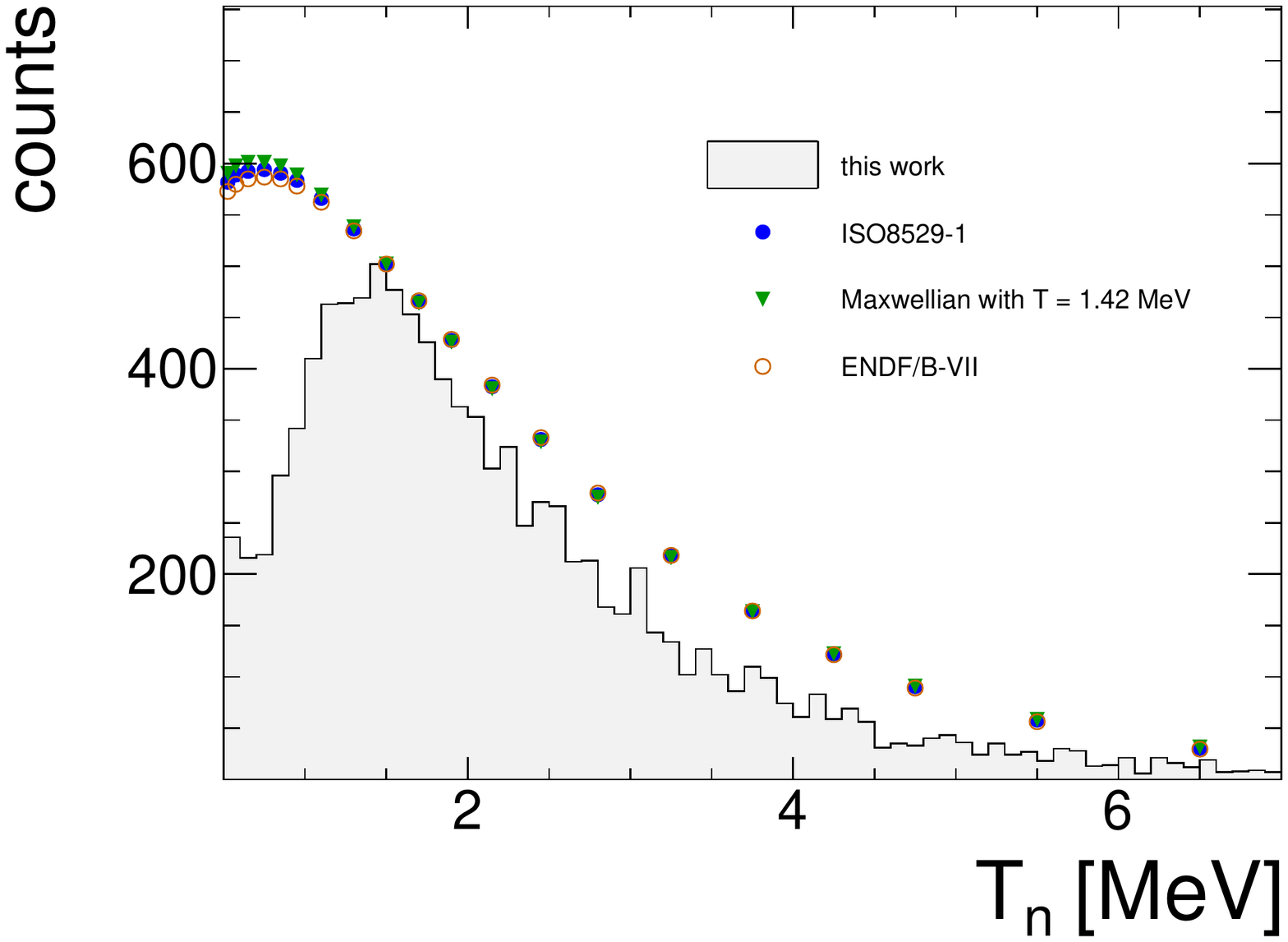}}
\caption{\label{figure:Ekin}
Fission-neutron kinetic-energy spectrum for $^{252}$Cf. The grey histogram 
is measured data. Also shown are the ISO~8529-1 recommendation (filled blue 
circles), a Maxwellian approximation (green triangles), and the ENDF/B-VII 
suggestion (open red circles) for the fission-neutron spectrum.
(For interpretation of the references to color in this figure
caption, the reader is referred to the web version of this
article.)
}
\end{center}
\end{figure}

Figure~\ref{figure:Ekin} shows the fission-neutron TOF spectrum from
Fig.~\ref{figure:TOF} converted to a neutron kinetic-energy spectrum. To 
convert from TOF to neutron kinetic energy, we used the 5~ns position of the 
gamma-flash shown in Fig.~\ref{figure:TOF} and the 106~cm distance between 
the $^{252}$Cf source and the center of the NE-213 liquid-scintillator cell.
The data were then rebinned linearly in kinetic energy. Also shown are three 
representations of the neutron-kinetic energy distribution constructed using 
the information presented by Thomas in Ref.~\cite{thomas14}. We note that 
there are small differences between the representations, so a normalization 
factor has been applied to each so that they coincide with our data at 
1.5~MeV. It should be emphasized that the present data have not been 
corrected for neutron-detection efficiency and that, with the present 
hardware neutron-detector threshold, neutron energies below 1~MeV cannot 
be corrected for reliably. At 1~MeV, the agreement between all three 
representations is essentially exact. Between 1~MeV and 4~MeV, the 
Maxwellian approximation and the ISO~8529-1 representation both lie below 
the ENDF/B-VII suggestion by 2\% and 1\%, respectively. By 5~MeV, the 
agreement between all three is again essentially exact. Above 6~MeV, a 
divergence between the representations begins and by 8~MeV, both the 
Maxwellian and the ISO~8529-1 representations lie above the ENDF/B-VII 
suggestion by about 9\% and 4\%, respectively. The $\pm$1\% level of 
agreement between all three representations of the $^{252}$Cf fission-neutron 
energy spectrum over the 
energy region from 1 to 6~MeV is well within any systematic uncertainty 
that we are likely to obtain in measurements of neutron-detection efficiency 
using the tagging technique presented here. Thus, they provide an excellent 
benchmark from which it will be possible to evaluate the neutron-detection 
efficiency.

\section{Summary}
\label{section:summary}

As a first step towards the development of an apparatus for the measurement 
of neutron-detection efficiency at our source-based fast-neutron irradiation 
facility, we have employed coincidence and time-of-flight measurement 
techniques to ``tag" neutrons emitted by a $^{252}$Cf source. The 
spontaneous-fission fragments are detected in a gaseous $^{4}$He scintillator 
detector. The neutrons are detected in a NE-213 liquid-scintillator detector. 
The resulting continuous polychromatic beam of tagged neutrons has a measured 
energy dependence that agrees qualitatively with expectations. This 
preliminary study strongly suggests that the method of neutron-energy tagging 
will work well and future investigations will concentrate on quantifying 
systematic effects in order to optimize the performance. We anticipate that 
the technique will provide a cost effective means for the characterization of 
neutron-detector efficiency, and note that this technique will work equally 
well for all spontaneous-fission neutron sources.

\section*{Acknowledgements}
\label{acknowledgements}

We acknowledge the support of the UK Science and Technology Facilities
Council (Grant nos. STFC 57071/1 and STFC 50727/1) and the European
Union Horizon 2020 BrightnESS Project, Proposal ID 676548.

\newpage
\bibliographystyle{elsarticle-num}

\end{document}